\documentclass[pre,twocolumn,showpacs,floatfix]{revtex4}

\usepackage{graphicx}

\begin{document}

\title{Molecular dynamics simulation of reversibly self-assembling shells
in solution using trapezoidal particles}

\author{D. C. Rapaport}

\email{rapaport@mail.biu.ac.il}

\affiliation{Department of Physics, Bar-Ilan University, Ramat-Gan 52900, Israel}

\date{November 19, 2012}

\begin{abstract}

The self-assembly of polyhedral shells, each constructed from 60 trapezoidal
particles, is simulated using molecular dynamics. The spatial organization
of the component particles in this shell is similar to the capsomer proteins
forming the capsid of a T=1 virus. Growth occurs in the presence of an
atomistic solvent and, under suitable conditions, achieves a high yield of
complete shells. The simulations provide details of the structure and
lifetime of the particle clusters that appear as intermediate states along
the growth pathway, and the nature of the transitions between them. In
certain respects the growth of size-60 shells from trapezoidal particles
resembles the growth of icosahedral shells from triangular particles studied
previously, with reversible bonding playing a major role in avoiding
incorrect assembly, although the details differ due to particle shape and
bond organization. The strong preference for maximal bonding exhibited by
the triangular particle clusters is also apparent for trapezoidal particles,
but this is now confined to early growth, and is less pronounced as shells
approach completion along a variety of pathways.

\end{abstract}


\maketitle

\section{Introduction}

Self-assembly at the molecular scale occurs in an environment where
thermal noise provides strong competition to the forces that drive growth;
in this respect such microscopic processes differ significantly from their
macroscopic counterparts. While direct experimental observation of the
details of supramolecular self-assembly is not readily achieved, computer
simulation, assuming the availability of simplified models capable of
capturing the essential details, ought to be able to supply information
that is otherwise inaccessible.

The formation of the capsid shells enclosing the genetic material of
spherical viruses \cite{cri56,cas62} is a well-known example of
self-assembly. The organization of capsid structures is simplified and the
construction specifications are minimal because the shells are assembled
from multiple copies of one or a small number of different capsomer
proteins \cite{bak99} and the structures satisfy icosahedral symmetry.
This information, however, provides little help in trying to determine the
assembly steps involved in forming the capsid. Even a highly simplified
version of the problem, in which capsomers spontaneously and reversibly
form complete shells under {\em in vitreo} conditions free of genetic
material \cite{ber87,pre93,cas04,zlo11}, remains opaque. The robustness of
self-assembly \cite{cas80} makes understanding the process in simplified
environments a worthwhile endeavor, especially since analogous processes,
inspired by the mechanisms employed by the virus itself, could provide a
basis for nanoscale chemical packaging with possible therapeutic uses
involving targeted delivery.

Molecular dynamics (MD) simulation \cite{rap04bk}, with its ability to
capture both the spatial and time-dependent properties of interacting
many-body systems, is capable of providing access to the shell assembly
pathways themselves and predicting the varying populations of partially
complete structures; this provides, in principle, a direct link with
experiment \cite{end02}. A simplified capsomer particle for use with MD
can be represented by a set of soft spheres rigidly arranged to produce an
effective molecular shape consistent with packing into a closed shell,
together with a set of interaction sites where attractive forces between
particles act. Reduced descriptions of this kind are designed to mimic the
relevant features of real capsomers that consist of folded proteins --
large molecules whose exposed surfaces have relatively complex landscapes
that are able to fit together to form the closed, strongly bound capsids.

The initial MD studies of this problem \cite{rap99,rap04a} were severely
restricted by limited computational resources and consequently focused on
demonstrating the feasibility of assembly in the absence of solvent,
subject to the restriction that the process was irreversible (meaning that
bonds, once formed, are unbreakable). Shells of size 60 were grown from
triangular and trapezoidal particles, the latter corresponding to the
structure of T=1 viruses, as well as shells of size 180 resembling T=3
viruses. This was followed by a more computationally demanding MD study of
reversible assembly (in which bonds break when sufficiently stretched) for
T=1 shells \cite{rap04a}, but while reversibility is more reasonable from
a physical perspective the approach required that smaller particle
clusters be decomposed at regular intervals to avoid kinetic traps due to
a lack of unbonded particles.

Increased computer performance permitted the inclusion of an explicit
atomistic solvent \cite{rap08,rap10a,rap10b} thereby eliminating the need
for enforced decomposition, but only for the case of triangular particles
assembling into 20-particle icosahedral shells. The explicit solvent
provides a means for collision-induced breakup of clusters without needing
them to come into direct contact; it also adds a diffusive component to
the otherwise ballistic particle motion, and serves as a heat bath for
absorbing and redistributing energy released when particles bond. These
simulations demonstrated that self-assembly proceeds via a sequence of
reversible stages, with a high yield of complete shells and a strong
preference for minimum-energy intermediate clusters. Though seemingly
paradoxical, reversibility provides the key to efficient self-assembly due
to its ability to prevent subassemblies becoming trapped in configurations
inconsistent with continued correct growth.

The goal of the present work is to extend the previous MD study of
icosahedral shell assembly in solution to the larger T=1 shells. Increased
shell size offers a broader range of growth possibilities, permitting
`entropic' effects to compete more strongly with the energetic preferences
dominating the growth of smaller shells. Comparing the outcomes of growth
simulations involving different shell sizes can provide insight into how
this factor influences growth and, in particular, which aspects of growth
observed previously are common to both the smaller and larger shells.

An alternative, even more simplified representation of capsomers can be
based on spherical particles, using directional interactions, and an
implicit solvent represented by stochastic forces \cite{hag06}. The
motivation for the present study, based on extended rather than spherical
particles, is that the capsomers are themselves extended bodies, with
complex shapes generally tailored to conform to the shells. Use of extended
particles means that the interaction range need not exceed the particle
size, allowing the design to be tuned to ensure that bonding forces are
maximized only when particles are correctly positioned and oriented, while
avoiding bond formation in other situations; this is reflected by the
absence of any incorrect growth in the simulations described here. Another
difference is in the solvent representation; the question of whether the
explicit solvent used here could be replaced by stochastic forces has not
been examined, although the former has the advantage that motions of
particles not in direct contact are correlated through the solvent, as would
be the case in a real fluid. In the case of block copolymers it has been
shown that self-assembly simulations based on implicit and explicit solvents
lead to very different outcomes \cite{spa11}; different solvent dynamics may
also help explain the fact that enhanced pentamer stability is observed, as
expected, when assembling triangular particles using an explicit solvent
\cite{rap08}, but not when the solvent is implicit \cite{mah12}.

MD simulations using complete all-atom descriptions of the capsomer proteins
\cite{fre06} are another possibility, but because of their complexity they
are presently limited to very short time intervals, adequate only for
examining preassembled shells. A further simplified MD approach involves
quasi-rigid bodies formed from hard spheres \cite{ngu07}. Monte Carlo
simulations have been used in assembly studies of particles of various
shapes \cite{che07,wil07,joh10}. A number of theoretical techniques for
studying capsid structure have been explored, including thin shells
\cite{lid03}, tiling \cite{twa04}, particles on spheres \cite{zan04},
stochastic kinetics \cite{hem06}, elastic networks \cite{hic06}, and
nucleation theory \cite{zan06}, as has a combinatorial approach to the
pathways \cite{moi10}. Focusing on the kinetic aspects of subassembly
concentration is another approach \cite{mor09,hag10} that is also used in
interpreting experimental results \cite{zlo99,van07} and analyzing the
effects of reversibility on growth \cite{zlo07}.

\section{Methods}

\begin{figure}
\begin{center}
\includegraphics[scale=0.38]{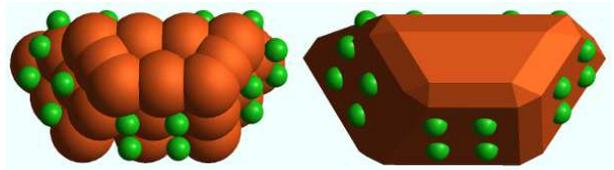}
\end{center}
\caption{\label{fig:01} (Color online) The component spheres and effective
shape of the trapezoidal particle; the small spheres denote the attraction
sites.}
\end{figure}

The two components in the MD system are the self-assembling model capsomer
particles and the solvent atoms. The particle, shown in Fig.\,\ref{fig:01},
features an extended, highly specific shape, together with multiple
attraction sites. It is formed from a rigid array of soft spheres arranged
to have the effective shape of a trapezoidal truncated pyramid, and 60
copies can be packed to make a closed shell; the design was introduced in
the earlier solvent-free study \cite{rap04a}. The lateral faces contain the
attractive interaction sites involved in bond formation and determine the
dihedral angles of the assembled shell; two of the adjacent short lateral
faces are perpendicular to the plane of the particle, allowing three
adjacent particles to form a planar triangular face, whereas the other two
faces are inclined to provide the required dihedral angle between adjacent
triangular shell faces.

Two kinds of interactions are used in the model \cite{rap10b}. Soft-sphere
repulsion is provided by the truncated Lennard-Jones potential,
\begin{equation}
u_s(r)=\left\{
\begin{array}{ll}
4 \epsilon [(\sigma/r)^{12}-(\sigma/r)^{6}+1/4] &
 \quad r<r_c=2^{1/6} \sigma \\[4pt]
0 & \quad r \ge r_c
\end{array}
\right.
\end{equation}
where $r$ is the separation, $r_c=2^{1/6} \sigma$ is the interaction cutoff,
with $\sigma$ approximating the effective sphere diameter, and $\epsilon$
determines the energy scale. Solvent atoms are represented using the same
interaction. In standard reduced MD units, $\sigma=1$ and $\epsilon=1$,
while both the solvent atoms and particle spheres have unit mass. The length
of the irregularly shaped particle (distance between the centroids of the
bonding sites in the opposite short faces) is 3.6 (MD units), the width
(between bonding sites in opposite long and short faces) 2.1, and the depth
(extent of top and bottom spheres) 2.7. In Fig.\,\ref{fig:01} the component
spheres are drawn with unit diameter.

The attractive interaction responsible for assembly consists of two parts
that blend together smoothly, a short-range, finite-depth harmonic well and
a medium range, inverse-power attraction,
\begin{equation}
u_a(r)=\left\{
\begin{array}{ll}
e (1/r_a^2+r^2/r_h^4-2/r_h^2) & \quad r<r_h \\[4pt]
e (1/r_a^2-1/r^2) & \quad r_h \le r<r_a
\end{array}
\right.
\end{equation}
Attraction acts selectively and occurs only between those sites in face
pairs that would be adjacent in a correctly assembled shell and, of these,
only between correspondingly positioned sites. Site pairs in the bound state
tend to lie near the bottom of the well ($r = 0$), but there is nothing to
prevent escape if sufficiently excited. The attraction changes form at the
crossover distance $r_h=0.3$, and ceases entirely at the cutoff $r_a=3$.
Individual pair interactions have no directional dependence, but when acting
together they contribute to correct particle positioning and orientation.
The only parameter varied in $u_a(r)$ is the overall attraction strength
$e$.

Standard MD methods \cite{rap04bk} are used for the simulations. The force
calculations employ neighbor lists for efficiency, with separate lists used
for the soft-sphere repulsive forces and the longer-range attractions. Once
all the forces acting on the soft spheres and attraction sites of the
particles have been evaluated they are combined to produce the total forces
and torques required for the translational and rotational equations of
motion; these are solved using leapfrog integration, with a time step of
0.005 (MD units). Constant-temperature MD is used to prevent heating due to
exothermic bond formation. The boundaries of the simulation region are
periodic and the region size is determined by the overall number density.
Preparation of the initial state and other details appear in \cite{rap08}.

Methods for cluster analysis were described previously \cite{rap10a}. For
any two particles, if each of their four corresponding attraction-site pairs
are closer than $r_b$ then the particles are considered bonded; setting
$r_b$=0.5 leads to quantitative results consistent with direct observation,
namely no spurious bond breakage or inappropriate bonds. The bond count of a
cluster, used in the analysis below, is the total number of bonded particle
pairs.

\section{Results}

\begingroup
\squeezetable
\begin{table}
\caption{\label{tab:1} Final cluster distributions for different interaction
strengths, $e$, expressed as mass fractions and grouped by cluster size into
monomers, clusters in different size ranges, and complete shells; the
fractions with the majority populations are shown in bold and the run
lengths are included.}
\begin{ruledtabular}
\begin{tabular}{cr|cccccc}
$e$ & Time steps & \multicolumn{6}{c}{Cluster mass fraction}		    		\\
    &       & Size: 1	& 2--10 & 11--30& 31--50    & 51--59	&     60    		\\
\hline
0.080 &$72 \times 10^6$&{\bf 0.997}& 0.003 & 0.    &	 0.    &     0.     &	  0.	\\
0.085 &$256\times 10^6$&{\bf 0.628}& 0.001 & 0.    &	 0.    &     0.     &	  0.371 \\
0.090 &$251\times 10^6$&     0.175 & 0.    & 0.    &	 0.017 &     0.022  &{\bf 0.786}\\
0.095 &$146\times 10^6$&     0.019 & 0.    & 0.039 &	 0.256 &{\bf 0.642} &	  0.044 \\  
0.100 &$149\times 10^6$&     0.008 & 0.002 & 0.085 &{\bf 0.473}&     0.432  &	  0.	\\  
\end{tabular}
\end{ruledtabular}
\end{table}
\endgroup

\begin{figure}
\begin{center}
\includegraphics[scale=0.21]{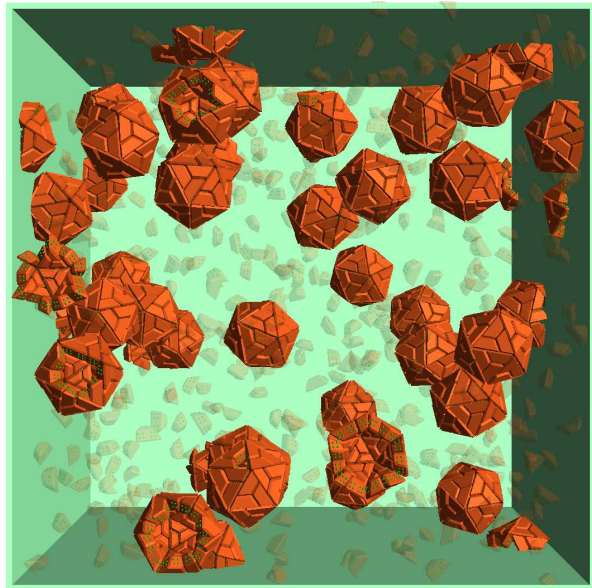}
\end{center}
\caption{\label{fig:02} (Color online) Late state of the $e$=0.090 run, with
the solvent omitted and particles not in complete shells shown
semitransparently; complete shells that cross periodic boundaries appear
open, an artifact of the visualization.}
\end{figure}

\subsection{Shell production}

The present simulations consider systems in which the total number of
trapezoidal particles and solvent atoms is 125\,000, contained in a cubic
region with an overall number density of 0.1; the particle concentration is
2.2\% (by number -- corresponding to a volume fraction of 0.045), enough, in
principle, for 45 complete shells. The runs cover a series of interaction
strength values, $e$, resulting in a variety of outcomes. Thermostatting
maintains a constant temperature of 0.667, equivalent to unit average
translational kinetic energy per particle or solvent atom; the corresponding
total energy drops as bonding occurs, e.g., for $e$=0.090 it falls from 1 to
0.7 over the course of the run.

The final cluster distributions and run lengths are summarized in
Table~\ref{tab:1}. For increasing $e$, over a relatively narrow range, these
vary from essentially no growth, through various yields of complete shells,
to cases in which there is abundant growth but no full shells. The values
for $e$=0.090 correspond to 36 complete shells, amounting to an 80\% yield.
In this run, and for $e$=0.085 where there is also significant shell
production, the almost complete absence of intermediate size structures when
growth ends is especially notable. The highest shell yield for trapezoidal
particles is achieved at $e$ approximately $0.6 \times$ the corresponding
triangular value \cite{rap08}, so that the overall binding energies per
particle in the two cases are similar.

Fig.\,\ref{fig:02} shows the $e$=0.090 system once shell growth is
practically complete. The fact that there is ample space for shell growth
without crowding is apparent. The mean separation of bound attraction sites
is only 0.024 (MD units). Complete shells are likely to enclose solvent
atoms since there are no interactions that prevent this.

\begin{figure}
\begin{center}
\includegraphics[scale=0.55]{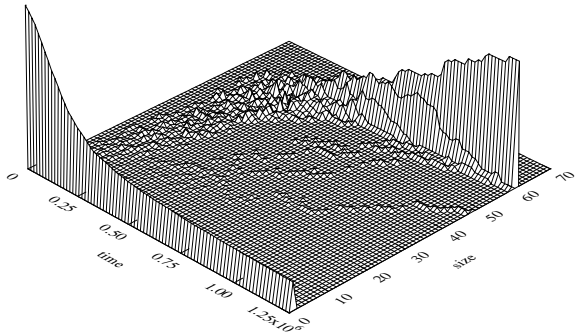}
\end{center}
\caption{\label{fig:03} Time-dependent cluster size distribution (mass
fraction) for $e$=0.090; the final peaks correspond to monomers and complete
shells; each grid interval along the time axis (MD units) corresponds to
$\sim 3 \times 10^6$ time steps.}
\end{figure}

\begin{figure}
\begin{center}
\includegraphics[scale=0.75]{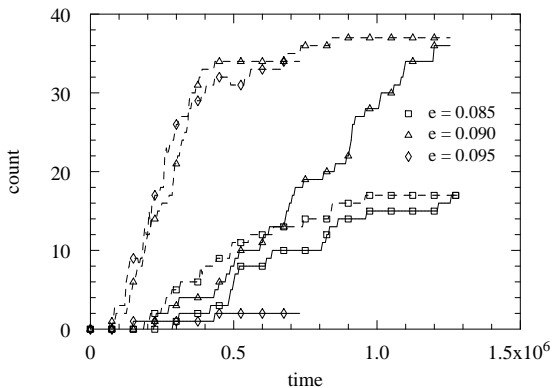}
\end{center}
\caption{\label{fig:04} Number of complete shells (solid lines) and combined
number of complete shells and large (size $> 50$) clusters (dashed lines) as
functions of time.}
\end{figure}

\subsection{Cluster size distributions}

The time-dependent cluster size distributions exhibit the same
$e$-dependence noted for triangular particles \cite{rap08}.
Fig.\,\ref{fig:03} shows the results for $e$=0.090; although appearing
similar to the icosahedral results, the time and size scales are
considerably larger. Two prominent features are the sharp bimodality of the
distribution and the absence of significant populations of intermediate size
clusters. Since clusters commence growing at different times there is no
evidence for coordinated growth \cite{mor09}.

Cluster growth rates are sensitive to $e$; the time-dependence of the number
of complete shells and the combined number of complete shells and large
clusters with size $> 50$ are shown in Fig.\,\ref{fig:04}. For $e$=0.085 and
0.090, the convergence of the cluster numbers towards the end of the runs,
irrespective of the different yields, reflects the fact that almost all
large clusters grow to completion. Entirely different behavior occurs for
$e$=0.095, where there are many large but incomplete clusters.

\subsection{Bond distributions}

\begin{figure}
\begin{center}
\includegraphics[scale=0.75]{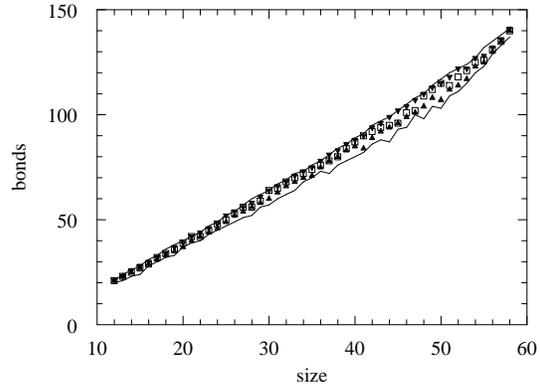}
\end{center}
\caption{\label{fig:05} Bond counts for different cluster sizes ($e$=0.090);
solid lines show the minimum and maximum observed counts, triangles the
ranges of counts accounting for $>$ 80\% of cases, and squares the most
frequent counts.}
\end{figure}

A simple way of classifying intermediate structures is based on the bond
counts defined earlier. Fig.\,\ref{fig:05} shows the measured variation in
bond count for each cluster size over the range of sizes where this is
significant ($e$=0.090). The results include the minimum and maximum bond
counts, the ranges of counts accounting for over 80\% of cases -- these
generally either include the maximum counts or lie just 1 or 2 below them --
and the most frequent counts. Bond count depends only weakly on $e$; the
average count (over all cluster sizes) for, e.g., $e$=0.1 is smaller by
approximately 0.9 (1\%), reflecting fewer breakup events that could increase
the fraction of more highly bonded clusters.

In the case of smaller clusters, the results are similar to icosahedra
\cite{rap08}, namely a strong preference for maximum bond counts, with over
90\% of the clusters below size 12 in this category. This effect is less
pronounced for larger clusters. Furthermore, unlike icosahedral shells,
structures formed by trapezoidal particles during the later stages of
assembly are sufficiently large to allow multiple, well-separated zones
where growth occurs independently (see Fig.\,\ref{fig:10} below).

\begin{figure}
\begin{center}
\includegraphics[scale=0.75]{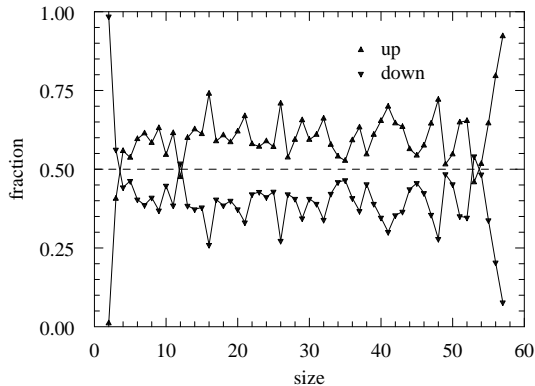}
\end{center}
\caption{\label{fig:06} Event fractions corresponding to cluster size
changes ($e$=0.090).}
\end{figure}

\begin{figure}
\begin{center}
\includegraphics[scale=0.75]{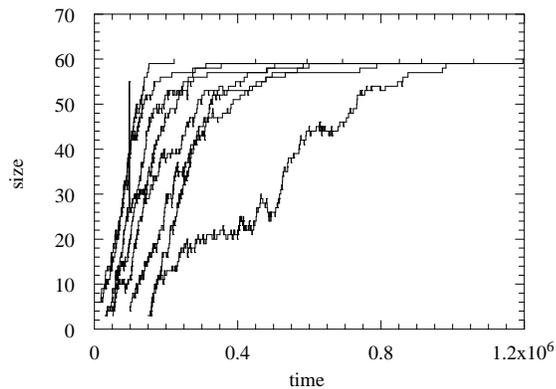}
\end{center}
\caption{\label{fig:07} Growth histories of several shells; the large spike
for one of the shells corresponds to a temporary merger of two big
clusters.}
\end{figure}

\begin{figure}
\begin{center}
\includegraphics[scale=0.75]{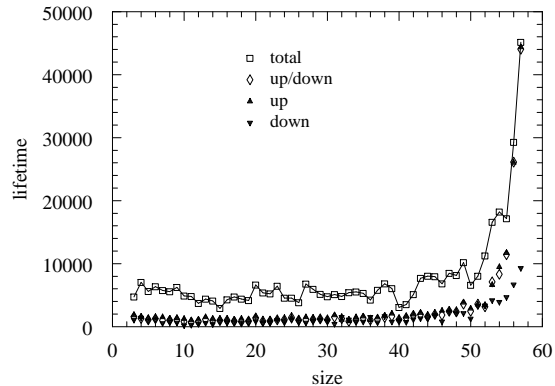}
\end{center}
\caption{\label{fig:08} Total and intermittent cluster lifetimes (MD units),
the latter also subdivided according to whether, in the subsequent event,
the size change is up or down ($e$=0.090).}
\end{figure}

\subsection{Reversible bonding}

Growth proceeds by means of a sequence of size-change events.
Fig.\,\ref{fig:06} shows the fraction of events experienced by clusters of
each size that correspond to up (growth) and down (breakup) size changes.
Practically all size changes are of unit magnitude (details not shown). 
With the notable exception of dimers, and to a lesser extent trimers, growth
is almost always more likely than breakup. Reversibility is important, but,
unlike the triangular particles \cite{rap08,rap10b} where the preferred
size-change direction varies strongly with cluster size, for trapezoidal
particles the preference for specific intermediate cluster sizes is reduced.
The effect of reversibility on assembly is apparent in the growth histories
shown in Fig.\,\ref{fig:07}, where size fluctuations are prominent.

Table~\ref{tab:2} shows the $e$-dependence of $P_g$, the probability of the
next event being growth, and $T_i$, the average intermittent lifetime (the
elapsed time between consecutive size-changing events) for the smallest
clusters. The extremely low dimer $P_g$ ($\sim$ 1\%) implies that
practically all dimer events amount to disappearance. Trimers are more
stable than dimers, as reflected in the reduced breakup probability
($=1-P_g$) and a $T_i$ value over 20$\times$ larger; in contrast, for
triangular particles \cite{rap08} the earliest appearance of enhanced
stability occurs for pentamers. The mean $P_g$ values for larger sizes are
included; for the 5--20 size range $P_g$ increases with $e$ as before, but
for 21--50 the trend is unclear because falling monomer availability also
affects the behavior.

\begingroup
\squeezetable
\begin{table}
\caption{\label{tab:2} Average growth probabilities, $P_g$, and intermittent
lifetimes (MD units), $T_i$, of the smallest clusters, for different $e$;
mean $P_g$ values for larger clusters are also shown.}
\begin{ruledtabular}
\begin{tabular}{cccc}
Size   &  $e$   & $P_g$ & $T_i$ \\
\hline
2      &  0.080 & 0.006 &\ \ 39 \\
       &  0.085 & 0.007 &\ \ 49 \\
       &  0.090 & 0.014 &\ \ 62 \\
       &  0.095 & 0.015 &\ \ 79 \\
\hline
3      &  0.080 & 0.193 &\ 790  \\
       &  0.085 & 0.284 & 1106  \\
       &  0.090 & 0.446 & 1684  \\
       &  0.095 & 0.516 & 1951  \\
\hline
4      &  0.080 & 0.261 & 1093  \\
       &  0.085 & 0.412 & 1580  \\
       &  0.090 & 0.554 & 1813  \\
       &  0.095 & 0.712 & 2223  \\
\hline
5--20  &  0.085 & 0.538 &  \\
       &  0.090 & 0.599 &  \\
       &  0.095 & 0.642 &  \\
\hline
21--50 &  0.085 & 0.601 &  \\
       &  0.090 & 0.609 &  \\
       &  0.095 & 0.586 &  \\
\end{tabular}
\end{ruledtabular}
\end{table}
\endgroup

Fig.\,\ref{fig:08} shows several kinds of cluster lifetime measurements,
namely the intermittent lifetime $T_i$, which is also subdivided according
to whether the subsequent event is a size increase or decrease, and the
total time a cluster exists at a given size $T_t$ (the sum over $T_i$). The
value of $T_i$ is based on all clusters appearing during the run, while
$T_t$ is obtained by tracking those clusters that correspond to the complete
shells and other large subassemblies present at the end. (For the final two
assembly stages, $T_t= 8.8 \times 10^4$ and $1.7 \times 10^5$.) Comparison
with triangular particles \cite{rap10a} shows reduced variability in $T_t$
at intermediate sizes. The ratio of $T_t$ to $T_i$ is an estimate of the
number of occasions a reversibly growing cluster reaches a particular size;
for most sizes this typically happens several (3--6) times.

\begin{figure}
\begin{center}
\includegraphics[scale=0.28]{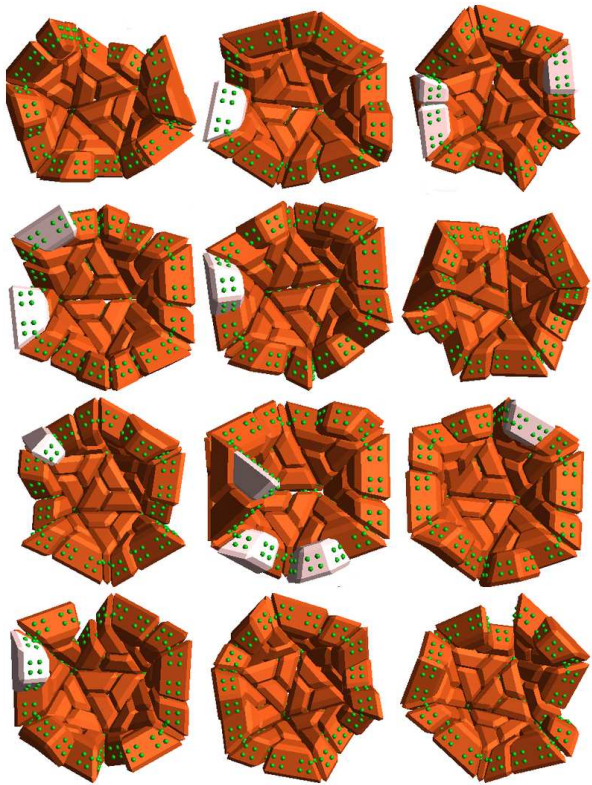}
\end{center}
\caption{\label{fig:09} (Color online) Clusters of size 30, oriented to show
their perimeters; the majority of the particles are also present in the
final shells into which the clusters develop, with the few that escape shown
in a lighter color/shade.}
\end{figure}

\begin{figure}
\begin{center}
\includegraphics[scale=0.28]{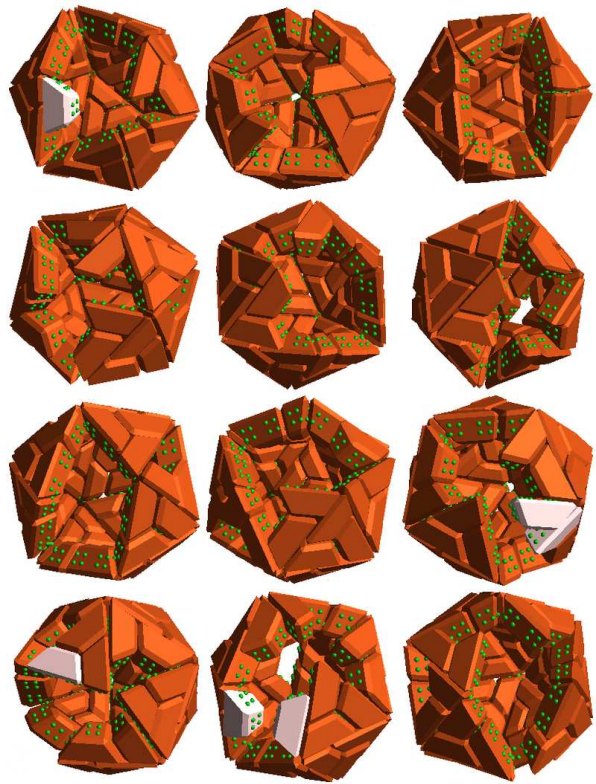}
\end{center}
\caption{\label{fig:10} (Color online) Clusters of size 50, oriented to show
the variation in hole number and shape.}
\end{figure}

\subsection{Visualizing structure and growth}

Examination of the intermediate clusters reveals considerable variation in
morphology not evident from the bond counts alone. A montage of 30-particle
clusters, each recorded the moment it first reached this size, is shown in
Fig.\,\ref{fig:09}. The perimeters have a variety of profiles with different
degrees of roughness, and the number of bonds observed in clusters of this
size ranges from 57 to 64 (80\% have 60--64 bonds). None of the clusters
have holes, although deep boundary indentations are potential precursors.
Fig.\,\ref{fig:10} shows a selection of incomplete shells containing 50
particles. The opportunity for independent growth in separate zones of the
structure is increased relative to icosahedral shells, and bond counts vary
between 103 and 117 (80\% have 107--115 bonds). The results are for
$e$=0.090, but the other $e$ values are similar. These incomplete shells do
not resemble the neatly truncated spheres employed in theoretical analysis
\cite{ber87,zan06}.

\begin{figure*}
\begin{center}
\includegraphics[scale=0.30]{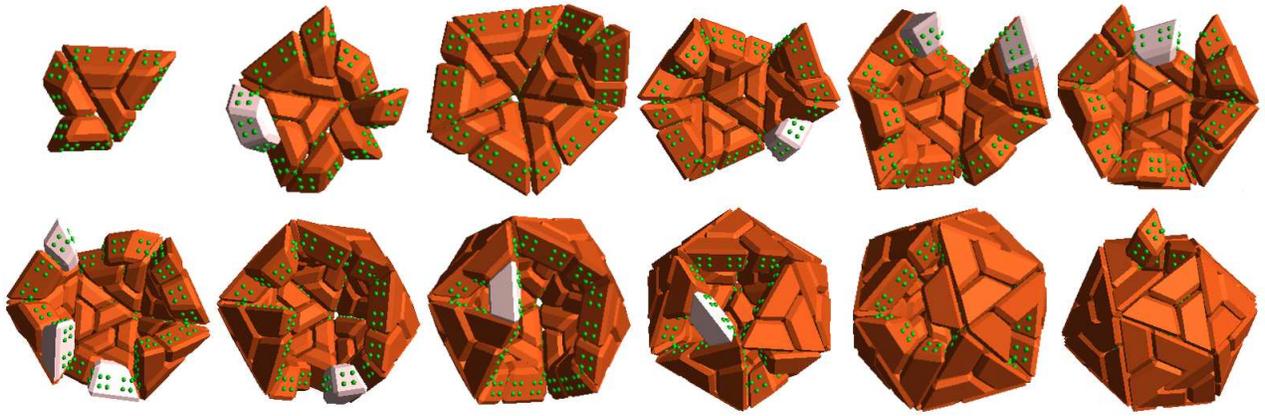}
\end{center}
\caption{\label{fig:11} (Color online) An example of shell growth (color
coding as in Fig.\,\ref{fig:09}); in the final image the shell is about to
close.}
\end{figure*}

\begin{figure}
\begin{center}
\includegraphics[scale=0.38]{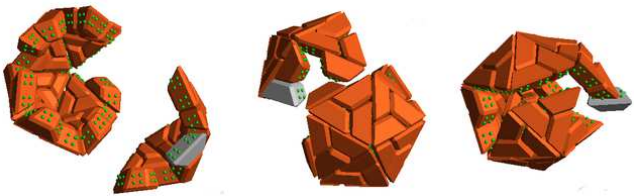}
\end{center}
\caption{\label{fig:12} (Color online) Stages in the successful merging of
two clusters; the last image shows the state an instant before final
bonding.}
\end{figure}

The image sequence in Fig.\,\ref{fig:11} shows several stages in the growth
of one of the $e$=0.090 shells. Here, as the shell begins to close, the
single large opening becomes several smaller holes that eventually fill.
Although growth by merging of extended clusters is a rare event, it is
occasionally observed; Fig.\,\ref{fig:12} shows an example in which clusters
of size 32 and 15 join. Not all such mergers persist, however, and the spike
in Fig.\,\ref{fig:07} corresponds to a shortlived merger.

\section{Conclusion}

The 60-particle shells that self-assemble from trapezoidal particles
considered in the present work share some of the previously observed growth
characteristics of icosahedral shells. All steps in the assembly process,
except at the very end, show strong reversibility, a characteristic of
systems only weakly out of equilibrium. Reversible bonding has a major
influence on shell production, by ensuring an adequate monomer supply and
allowing error correction to avoid incorrect structures. There is a clear
preference for the most highly bonded clusters during early growth, but
while this effect persists throughout the growth of icosahedral shells, it
is less prominent here. In both cases growth is rate-limited by dimer
formation; however, the different particle shapes lead to changes in the
intermediate cluster properties. The larger shells considered here offer
more opportunity for independent growth in well-separated zones of the
partial structures. Although the present focus is on the self-assembly
dynamics of polyhedral shells, key aspects of the observed behavior ought to
be relevant for other kinds of microscopic assembly phenomena.

\bibliography{bigassem}

\end{document}